\documentclass[aps,prx,twocolumn,floatfix,nofootinbib,superscriptaddress, longbibliography]{revtex4-2}
\pdfoutput=1
\usepackage{amsmath}
\usepackage{graphicx}
\usepackage{amssymb}
\usepackage{bbm}

\usepackage{braket}
\newcommand{\nuc}[2]{\ensuremath{^{#1}\textrm{#2}}}
\newcommand{\I}{{\mathrm i}}
\newcommand{\rd}{\mathrm d}
\newcommand{\ham}{{\mathcal H}}

\begin{document}
\title{Transforming pure and mixed states using an NMR quantum homogeniser}
\author{Maria Violaris}
\thanks{These two authors contributed equally}
\affiliation{Clarendon Laboratory, University of Oxford, Parks Road, Oxford OX1 3PU, United Kingdom}

\author{Gaurav Bhole}
\thanks{These two authors contributed equally}
\affiliation{Clarendon Laboratory, University of Oxford, Parks Road, Oxford OX1 3PU, United Kingdom}

\author{Jonathan A. Jones}
\affiliation{Clarendon Laboratory, University of Oxford, Parks Road, Oxford OX1 3PU, United Kingdom}

\author{Vlatko Vedral}
\affiliation{Clarendon Laboratory, University of Oxford, Parks Road, Oxford OX1 3PU, United Kingdom}
\affiliation{Centre for Quantum Technologies, National University of Singapore, 3 Science Drive 2, Singapore 117543}

\author{Chiara Marletto}
\affiliation{Clarendon Laboratory, University of Oxford, Parks Road, Oxford OX1 3PU, United Kingdom}

\date{\today}

\begin{abstract}
The universal quantum homogeniser can transform a qubit from any state to any other state with arbitrary accuracy, using only unitary transformations to perform this task. Here we present an implementation of a finite quantum homogeniser using nuclear magnetic resonance (NMR), with a four-qubit system. We compare the homogenisation of a mixed state to a pure state, and the reverse process. After accounting for the effects of decoherence in the system, we find the experimental results to be consistent with the theoretical symmetry in how the qubit states evolve in the two cases. We analyse the implications of this symmetry by interpreting the homogeniser as a physical implementation of pure state preparation and information scrambling.
\end{abstract}

\maketitle

\section{Introduction}

The quantum homogeniser can transform a qubit initialised in any state, whether pure or mixed, arbitrarily close to any other state. It was originally proposed as a theoretical model for analysing many-body entanglement within unitary thermalisation, with an additional possible application as a quantum safe \cite{ziman_quantum_2001, scarani_thermalizing_2002}. More generally, it can be used to implement processes such as quantum information scrambling and pure state preparation using only unitary interactions.  In this work we present a nuclear magnetic resonance (NMR) implementation of the quantum homogeniser, using a system of four qubits. We compare the limiting cases of transforming a qubit from a mixed to a pure state and the reverse process, and explain how the entropy changes in quantum homogenisation are consistent with unitary quantum theory.

The quantum homogeniser is a machine consisting of $N$ identical \textit{reservoir} qubits. These each interact, one by one, with the \textit{system} qubit (the qubit whose state is to be transformed) via a unitary \textit{partial swap}:
\begin{equation} \label{partial swap}
U = \cos\eta\,\mathbbm{1} + \text{i}\sin\eta\,S.
\end{equation}
The partial swap is a combination of the identity, $\mathbbm{1}$ (which does nothing to the two input qubits), and \textsc{swap} operation, $S$ (which swaps the states of the two input qubits), weighted by the coupling strength parameter $\eta$.

It has been shown \cite{ziman_quantum_2001} that if the system qubit interacts with $N$ reservoir qubits via the partial swap, then as $N \to \infty$, the system qubit state converges to the original state of the reservoir qubits, for \textit{any} coupling strength $\eta \neq 0$. Furthermore, all of the reservoir qubits after the interaction are within some distance $d$ of their original state, which can be made arbitrarily small as coupling strength $\eta \to 0$. The limit of a perfect homogenisation is achieved asymptotically, as $N$ tends to infinity, with an infinitesimal $\eta$. In that limit, any system qubit $\rho$ is sent to the reservoir qubit state $\xi$, with all the reservoir qubits remaining unchanged:
\begin{equation}
    U_N...U_1 (\rho\otimes \xi^{\otimes N})U^{\dagger}_1 ...U^{\dagger}_N\approx\xi^{\otimes N+1}
\end{equation}
where $U_k := U \otimes (\otimes_{j\neq k} \mathbbm{1}_j)$ denotes the interaction between the system qubit and the $k^{\text{th}}$ reservoir qubit. As already remarked in \cite{ziman_quantum_2001}, this is not in contrast with the no-cloning theorem, because the homogeniser realizes an approximate, not exact, copying of the quantum state $\xi$.

The information about the original system qubit's state is seemingly erased, despite all the interactions being unitary and thus information-preserving. The information has actually become stored in the infinitesimal entanglement between infinitely many reservoir qubits, which sums to a finite value \cite{ziman_quantum_2001}. This means the homogeniser can be considered as a unitary implementation of an eraser, in the limit of infinitely many qubits.

Erasure began its critical role in fundamental physics with Landauer's principle, used by Bennett to solve the Maxwell's Demon paradox \cite{bennett_thermodynamics_1982}. Since the logical reset process is key for classical information processing, the limits of erasure are likely to constrain practical devices in the near future \cite{zhirnov_limits_2003}. While commercial processors are several orders of magnitude away from the fundamental limits on erasure, experiments are beginning to reach the required sensitivity \cite{hong_experimental_2016, berut_experimental_2012, peterson_experimental_2016}. However, the physical implications of information erasure remain controversial \cite{lopez-suarez_sub-kbt_2016, hemmo_physics_2019}. There are various proposals for the true constraints on erasure when performed by a physical machine \cite{goold_nonequilibrium_2015, mohammady_minimising_2016, hanson_landauers_2017}. Experimental implementations of systems that perform erasure arbitrarily well through unitary interactions, such as the quantum homogeniser, could enable fundamental insights into the limits of information erasure within quantum theory.

\begin{figure*}[tb]
\centering
\includegraphics{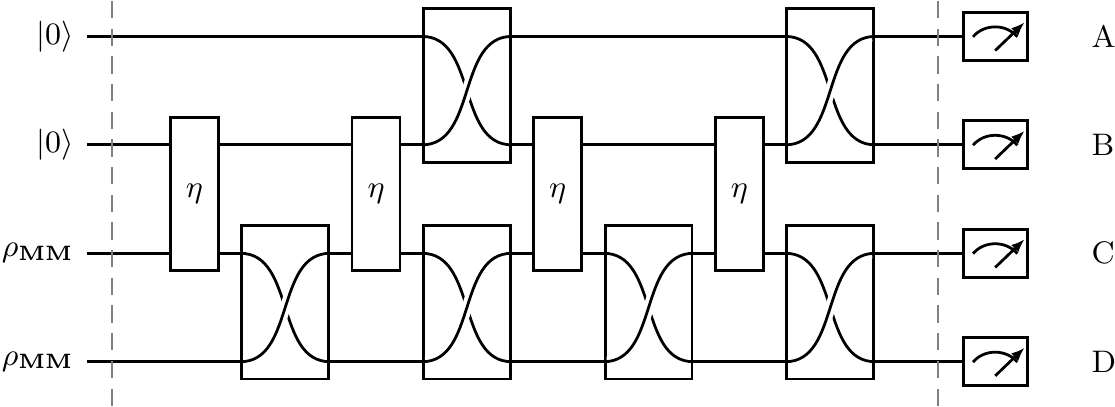}
\caption{A quantum homogeniser with two system qubits and two reservoir qubits designed to operate on a linear chain system where two qubit gates are possible only between adjacent qubits. Homogenisation is achieved using the partial swap gates, labelled as $\eta$, which connect the two middle qubits, while full \textsc{swap} gates are used to rotate the system and reservoir registers to bring other qubit pairs into contact. The dashed lines divide the simulation into its initialisation, homogenisation and readout phases. Qubit labels on the right hand side correspond to spins in the NMR spin system in Fig.~\ref{ham}.}\label{circuit}
\end{figure*}

In general, the homogeniser performs quantum information scrambling, where a local quantum state becomes distributed in many-body entanglement and correlations with other qubits \cite{lewis-swan_dynamics_2019}. Scrambling is a specific type of decoherence, which has been incorporated into unitary models for thermalisation in quantum thermodynamics \cite{cazalilla_focus_2010, bohrdt_scrambling_2017}. More exotically, it has also been proposed as a model for the information processing of a black hole \cite{hayden_black_2007, landsman_verified_2019}. Additionally, it provides a mechanism for cryptography: a scrambled qubit's original quantum state can only be recovered with the classical knowledge of its past interactions, securing the quantum information \cite{ziman_quantum_2001}. Homogenisation through scrambling forms a many-body entangled system, notoriously difficult to probe analytically due to the exponential complexity when entangling additional qubits. This heightens the need for experimental implementations of information scrambling systems.

As a special case of homogenisation, the homogeniser can be used to transform a qubit from any state to a pure state, to arbitrary accuracy. Pure state preparation has crucial practical implications for the resources required for quantum computation. The achievement of fault-tolerant quantum computation requires error correction, which has been proven to require a constant supply of pure ancillary qubits \cite{aharonov_fault-tolerant_2008}. Hence, quantum models that can produce pure states through unitary dynamics are of particular interest for quantum technology.

\section{Experimental Simulation}
The quantum circuit used in our simulation is shown in Fig.~\ref{circuit}. This circuit simulates the homogenisation between two system qubits, initially in the pure state $\ket{0}$, and two reservoir qubits initially in the maximally mixed state $\rho_{\rm MM}$, and has been designed for implementation on a linear chain where two-qubit gates are only available between adjacent qubits. Partial swap gates are implemented between qubits B and C, while full \textsc{swap}s are used to rotate the system and reservoir registers, permitting indirect contact between any pair of qubits.

As we are using an NMR implementation for our simulation we do not have access to qubits in pure states, but instead use ensemble qubits in \textit{pseudo-pure} state \cite{Cory1997,Cory1998a}, or \textit{effective-pure} states \cite{Gershenfeld1997,Knill1998}. As the pure component of these mixed states evolves in precisely the same way under unitary transformations as the desired pure state, and the maximally mixed component is not detectable in NMR experiments,  pseudo-pure states behave exactly like the corresponding pure states except that the signal intensity is reduced. Note, however, that preparation of pure initial NMR state is possible in special cases \cite{Anwar2004}, and that NMR implementations run using such states produce identical results to those with pseudo-pure states except for the increased signal size \cite{Anwar2004a,Anwar2004b}.

The four qubit state before the final readout stage depends on the coupling strength $\eta$, but in general is an entangled state. However the final readout stage involves an implicit partial trace over the other three qubits, leaving four separate single-qubit states, all of which lie along the $z$-axes of the respective Bloch spheres, permitting the state to be fully characterised by measurements in the computational basis. We can write for each qubit
\begin{equation} \label{general state}
\rho=\frac{\mathbbm{1}}{2}+f(\eta)\times\frac{\sigma_z}{2}
\end{equation}
where 
\begin{equation}\label{fZ}
 f={\rm tr}(\rho\sigma_z)   
\end{equation}
lies in the range $\pm 1$, and corresponds to the difference between the probabilities of finding $\ket{0}$ or $\ket{1}$ when measuring a qubit in the computational basis. For our circuit $f$ is confined to lie between 0 and 1, and in particular we find for the individual qubits the forms
\begin{equation} \label{form B}
f_B=\cos^4(\eta)
\end{equation}
and $f_C=1-f_B$, so that $f_B$ falls smoothly from 1 to 0, while $f_C$ rises in the opposite way. Similarly
\begin{equation} \label{form A}
f_A=4\cos^2(\eta)-9\cos^4(\eta)+8\cos^6(\eta)-2\cos^8(\eta)
\end{equation}
also falls from 1 to 0, but following a more complex pattern, with $f_D=1-f_A$ once again rising in the opposite way.

As the coupling strength $\eta$ is increased the homogenisation becomes more effective, and for circuits with the same number of system and reservoir qubits the states are completely interchanged in the limiting case $\eta=\pi/2$. This symmetry permits the roles of the system and reservoir qubits to be interchanged, and so this homogenisation circuit can equally well be viewed as a process that randomises the pure qubits or polarises the mixed qubits.

\subsection{The NMR system}
\begin{figure}[tb]
\centering
\includegraphics[width=90mm]{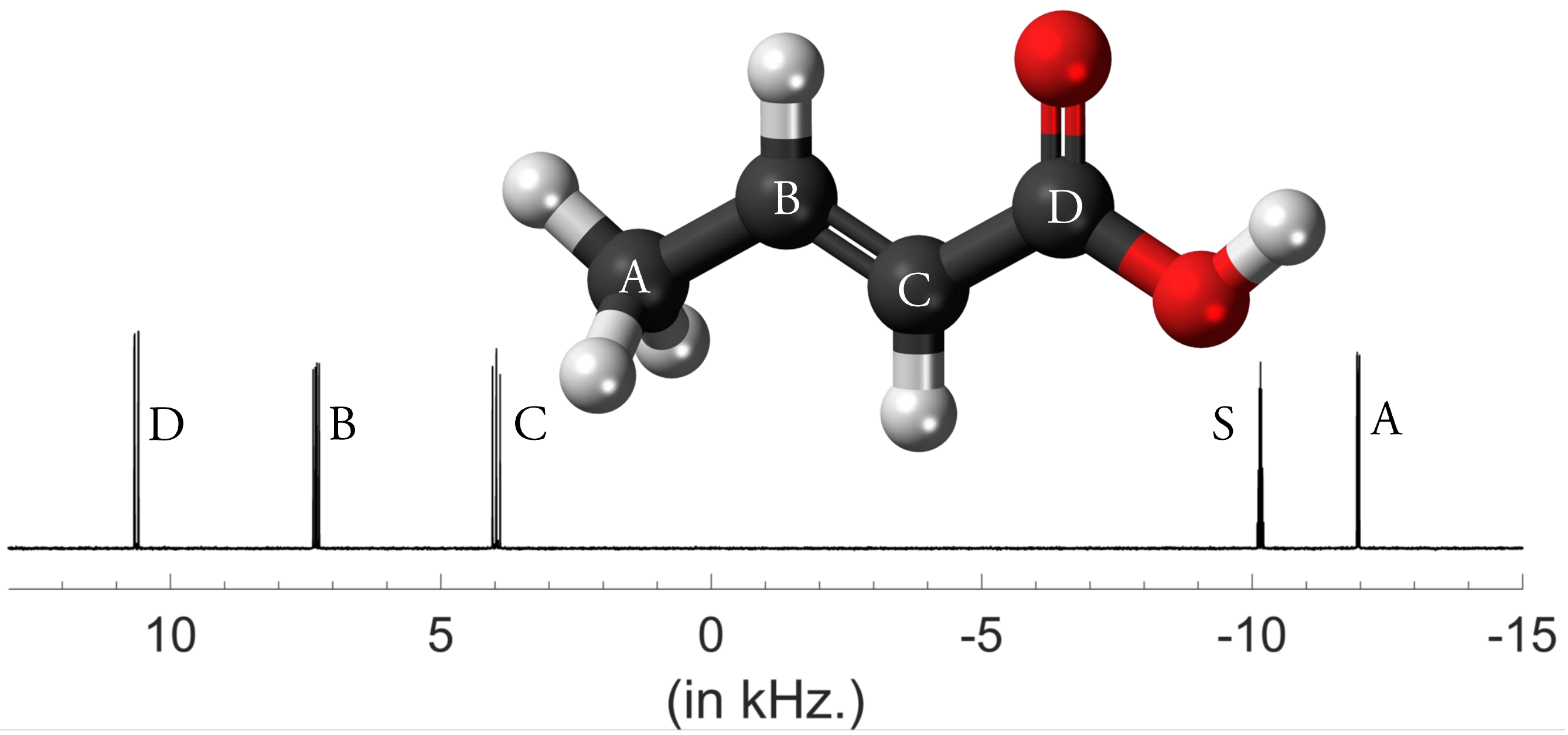}
\caption{The molecular structure and NMR spectrum for \nuc{13}{C} labelled crotonic acid with \nuc{1}{H} decoupling. The multiplet labelled S comes from the solvent, deuterated acetone; the second solvent peak at around 16.4\,kHz lies outside the spectral range plotted here. As in all NMR spectra the scale of the vertical axis is arbitrary and frequencies are measured from the RF transmitter frequency with frequency increasing from right to left. For more details see Appendix~\ref{NMR}.}
\label{ham}
\end{figure}

Our experiment to simulate a quantum homogeniser was performed on a four-qubit liquid-state NMR quantum processor, given by the four \nuc{13}{C} nuclei in a sample of fully \nuc{13}{C} labelled crotonic acid dissolved in deuterated acetone \cite{Boulant2002,Li2019}, as described in Appendix~\ref{NMR}. The Hamiltonian parameters obtained on a 600\,MHz (\nuc{1}{H} frequency) Varian Unity Inova spectrometer at 300\,K are listed in the Appendix and are the same as given in \cite{Bhole2020}. This four-qubit quantum processor can be approximated by a linear chain with strong nearest-neighbour couplings (between 42 and 72\,Hz) and weak long-range couplings (no more than 7\,Hz). A $^{13}\mathrm{C}$ NMR spectrum of the thermal equilibrium state with $^{1}\mathrm{H}$ decoupling is shown in  Fig.~\ref{ham}. The multiplet labelled S comes from the solvent, and so can be ignored.

\subsection{Initialisation}
NMR experiments \cite{EBWbook} are performed on macroscopic ensembles in high temperature thermal states. For this four-spin homonuclear system the thermal equilibrium state can be approximated as
\begin{equation}
\rho_\textrm{th}\approx\frac{\mathbbm{1}}{16}+p\left(A_z+B_z+C_z+D_z\right)
\end{equation}
where $p\sim10^{-6}$ is the thermal population difference and $A_z=(\sigma_z/2)\otimes\mathbbm{1}\otimes\mathbbm{1}\otimes\mathbbm{1}$ corresponds to the $z$ spin state of spin A, and similarly for the other three terms. The first term, which is the four-spin maximally mixed state, is not visible in NMR experiments, and the population difference simply determines the signal strength, so it is common to describe this state using just the \textit{deviation density matrix} \cite{Chuang1998b,Jones2011}, which is $A_z+B_z+C_z+D_z$.

As the thermal state is highly mixed NMR quantum information processing experiments normally begin with the preparation of a \textit{pseudo-pure} state \cite{Cory1997,Cory1998a}, or \textit{effective-pure} state \cite{Gershenfeld1997,Knill1998}, whose deviation density matrix corresponds to the desired pure state. Placing spins A and B in the pseudo-pure state $\ket{00}$ and spins C and D in the maximally mixed state corresponds to the deviation density matrix $A_z+B_z+2A_zB_z$ \cite{Jones2011}, but, as discussed below, only the first two terms give any visible signal in our experiments and so it suffices to use the simpler state $A_z+B_z$. This is trivial to prepare from the thermal state by applying $90^\circ$ excitation pulses to spins C and D followed by a crush gradient \cite{Jones2011}.

\subsection{Homogenisation}
The homogeniser circuit was implemented using gradient-ascent pulse engineering (GRAPE) \cite{Khaneja2005}, which assembles a single shaped pulse from a large number of short segments, each of which is individually controlled. We designed GRAPE pulses to implement each logic gate in Fig,~\ref{circuit}, except that when two \textsc{swap} gates are written vertically above one another a single GRAPE pulse was used to implement both gates together. Separate pulses were prepared for ten different partial swap gates, varying $\eta$ in $10^\circ$ steps between $0$ (an identity gate, for which the simplest implementation is just to omit the gate entirely) and $90^\circ$, corresponding to a full \textsc{swap}.

Although we have described our NMR spin system as a four-qubit device, the molecule also contains spin-$1/2$ \nuc{1}{H} nuclei, which are coupled to the four \nuc{13}{C} nuclei we use as qubits. (The \nuc{16}{O} nuclei are spin-$0$, and so can be safely ignored.) The conventional approach to this problem is to use continuous \nuc{1}{H} decoupling \cite{EBWbook} throughout the pulse sequence, but we were unable to obtain good results with this method as the length of the pulses limited the radio frequency (RF) power that could be used. Instead we repeated the approach used previously in this system in which we avoid decoupling during the pulse sequence, but designed each GRAPE pulse to tolerate couplings to \nuc{1}{H} nuclei \cite{Bhole2020} by averaging their fidelity over the variation in the \nuc{13}{C} Hamiltonian which arises from such coupling terms, as described in Appendix~\ref{Robust}. Pulses were also designed to tolerate some RF inhomogeneity by averaging their performance over a range of RF amplitudes.

When designing GRAPE pulses in NMR it is frequently necessary to add amplitude penalties to the function being optimised to prevent the algorithm from finding solutions with unfeasibly high RF powers. We adopted a simpler approach, using a single pre-determined amplitude for each pulse and varying only the RF phase between individual segments. In addition to sidestepping the need for amplitude limits this simplifies the underlying search, but at the cost of requiring that the individual segments are quite short, so that the effective rotation induced by each segment is small. Moreover, this approach to GRAPE avoids the computationally expensive operation of matrix exponentiation, leading to enormous speed-ups over conventional approaches. As a result this approach, when combined with the sub-system GRAPE approach described in \cite{Ryan2008}, could lead to efficient scalable control. Further details of this approach can be found in Appendix~\ref{PhaseOnly}.

\subsection{Readout}
The circuit in Fig.~\ref{circuit} assumes conventional projective measurements in the computational basis, so that it is necessary to repeat the experiment many times to estimate $f$, the $z$-component of the Bloch vector describing each qubit. Here the ensemble nature of NMR comes into its own, as NMR spectra directly reveal the desired expectation value.

Direct observation of the NMR signal, known as the free induction decay \cite{EBWbook}, reveals the expectation value of the $x$ and $y$ components of the Bloch vector. To observe the $z$-component we first apply a crush gradient, dephasing any pre-existing $xy$ components, and then apply a $90^\circ$ pulse to excite all four spins. The integrated signal intensity of each of the four multiplets seen in Fig.~\ref{ham} is then proportional to $f$. Note that integrating to find the total signal in each multiplet is equivalent to performing a partial trace over the other spins \cite{Cummins2002}.

Because we are integrating each multiplet it is not necessary to apply \nuc{1}{H} decoupling even during readout. However it is desirable to do so, as this reduces the width of each multiplet and so reduces the effects of noise in the integrated signal. To obtain accurate integrals it is important to process the data carefully, paying attention to phasing and baseline correction \cite{HSbook}. As NMR signal intensities are only \textit{proportional} to the desired $z$-component it is essential to obtain a suitable reference intensity against which all other intensities can be normalised. In the results below we use two different choices of normalisation, which emphasise different features of the experimental results. The use of normalised intensities means that experimental errors can take measured values of $f$ slightly outside the theoretical limits of $\pm 1$, and such apparently unphysical values should not cause concern.

\section{Results}
\begin{figure}[tb]
\centering
\includegraphics[width=85mm]{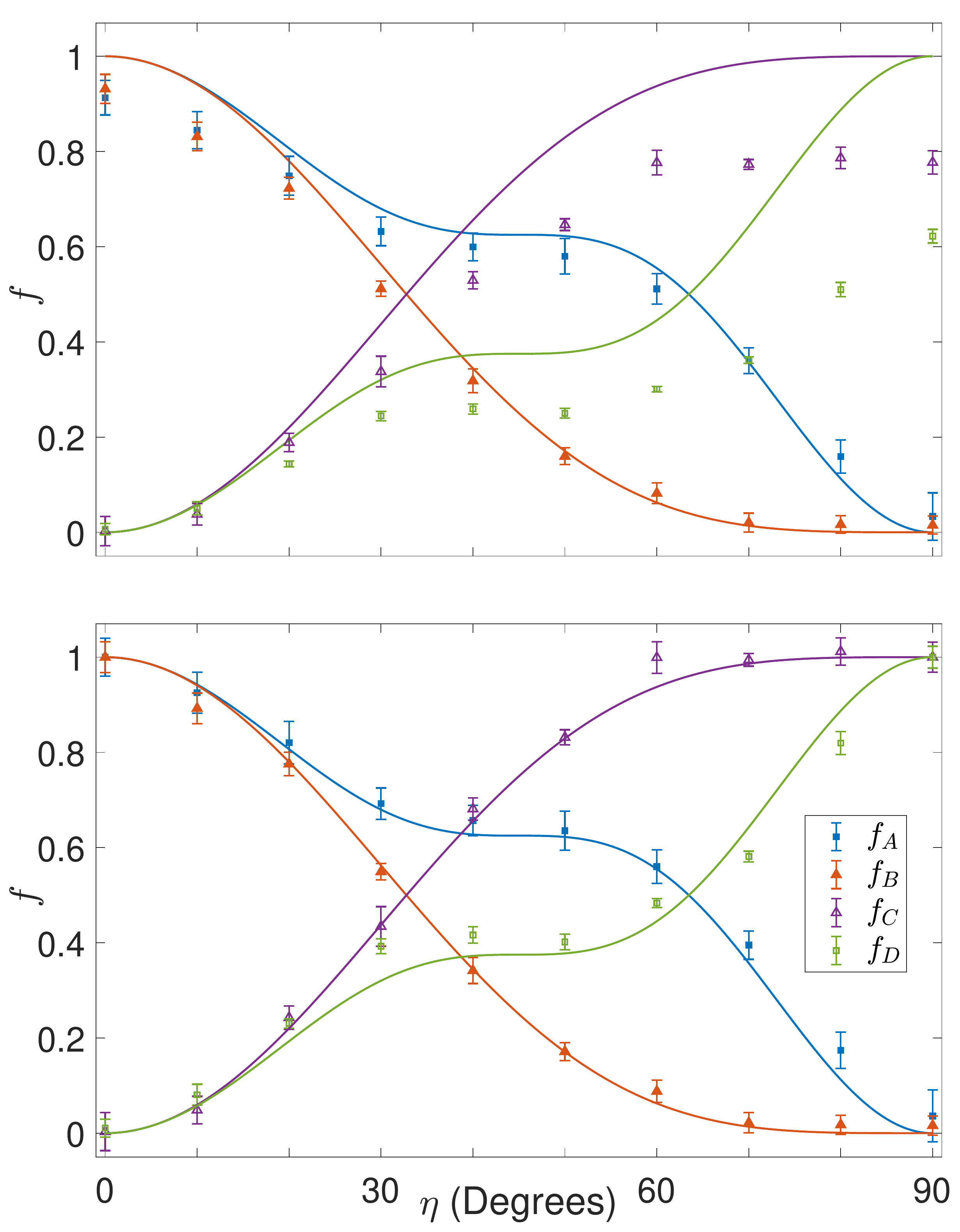}
\caption{Experimental results and theoretical predictions for the quantum homogeniser. Data-points show the results from ten repetitions of the experiment, while solid lines show theoretical calculations using Eqns. \ref{general state}--\ref{form A}. For the upper panel signal strengths were normalised against the average intensity from spins A and B in the initial state $A_z+B_z$, while in the lower panel the signal from each spin is normalised against a reference signal chosen for that spin.}
\label{experiment}
\end{figure}

The experimental results are shown in Fig.~\ref{experiment}, with the two sub-figures corresponding to the two different normalisation choices. In each case the lines show the expected polarisations for each of the four qubits calculated as described in the previous section, while data-points show the measured polarisations on the corresponding spins. Each experiment was repeated ten times, with the error bars showing the standard deviation around the mean.

The upper panel shows the results of normalising each intensity with respect to the average intensity of the A and B multiplets in the initial state. The experimental signal intensity is systematically lower than expected, with almost all data-points lying below the theoretical lines. The agreement for spins A and B is generally better than for spins C and D, although in all cases agreement is best close to $f=0$.

This general pattern of signal loss is easily explained as arising from the inevitable errors in any experimental implementation. Decoherence will normally lead to signal loss, and so will coherent errors which become effectively incoherent when averaged over the experimental ensemble, for example over different RF powers in different parts of the sample. Finally the readout process, including the initial crush gradient and the implicit partial trace, itself removes all terms in the density matrix other than single spin $z$-magnetisation. Thus errors of any kind can \textit{only} appear as a change in the measured values of $f$, usually reducing these towards zero.

The experimental asymmetry observed between the erasure of pure spins A and B and the polarisation of initially mixed spins C and D, in comparison with the symmetry of the theoretical predictions, is also easily understood. For spins A and B the signal loss acts in the same direction as the quantum homogeniser process, but for spins C and D it acts against the desired process, making the effects easier to see.

The lower panel shows the results when each spin was individually normalised against its intensity in the $\eta=0$ spectrum (for spins A and B) or $\eta=90^\circ$ spectrum (for spins C and D). This removes the effects of signal loss during the \textsc{swap} gates, and for C and D also removes losses due to partial swap gates. The experimental data-points now lie much closer to the theoretical predictions and the expected symmetry is largely restored.

\section{Interpretations}

The experimental data show the convergence of the system qubit to the state of the reservoir qubits. This has been tested for the limiting cases of transforming a qubit from a mixed state to a pure state and from a pure to a mixed state, demonstrating that the homogenisation is effective regardless of the initial states of the system or reservoir qubits. After accounting for the bias towards mixed states caused by decoherence, the experimental results are consistent with the theoretical symmetry in the evolution of how the states evolve, where the states for the pure-to-mixed homogenisation vary inversely compared to the states for the mixed-to-pure homogenisation.

\begin{figure}[tb]
\centering
\includegraphics[width=85mm]{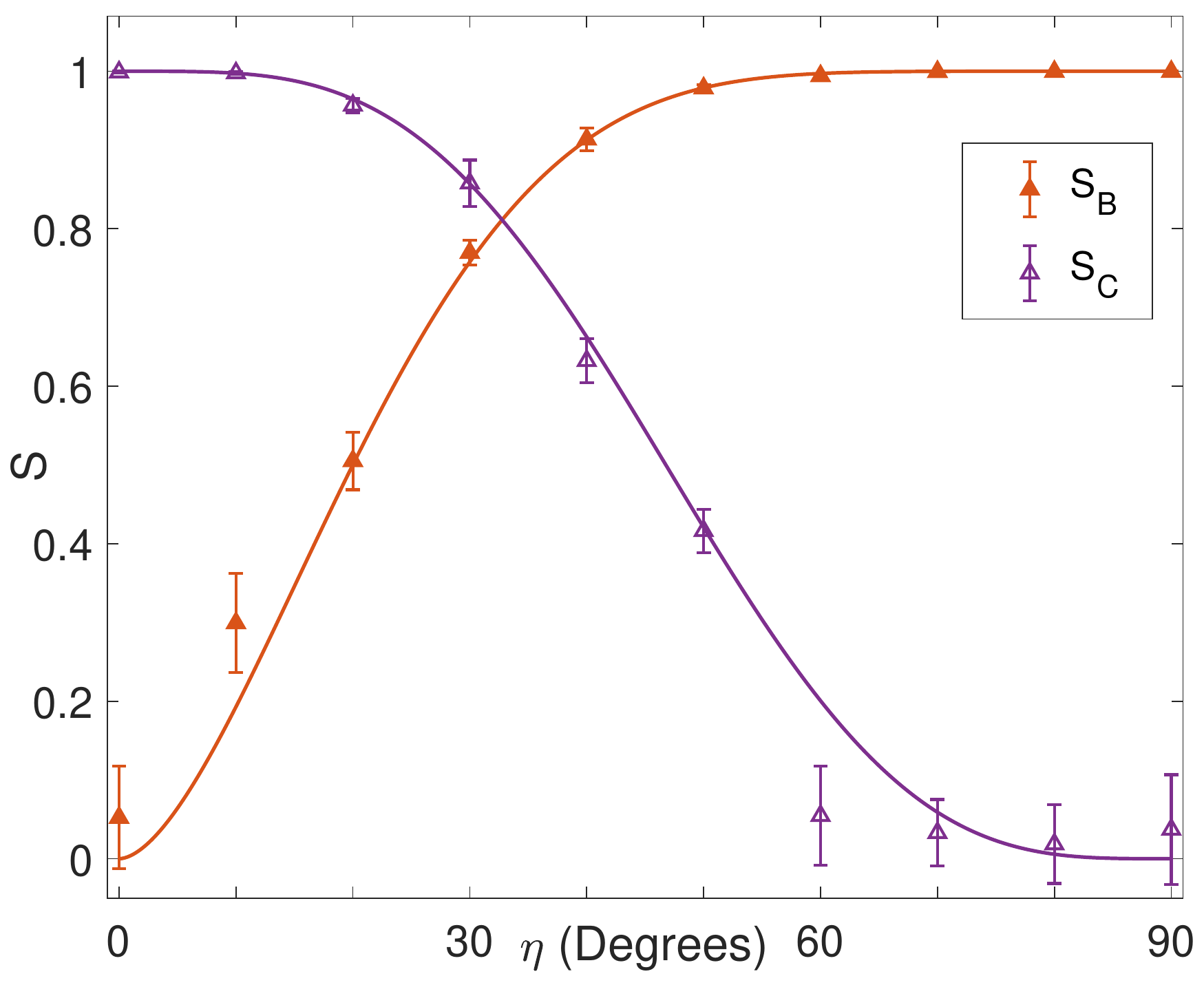}
\caption{Von Neumann entropies of qubits B and C against coupling strength, calculated from the experimental results and plotted against the theoretical predictions.}
\label{VN B and C}
\end{figure}

\paragraph*{Pure state preparation.---}
Figure \ref{VN B and C} plots the von Neumann entropy
\begin{equation}
S=-\left(\frac{1+f}{2}\right)\log_2 \left(\frac{1+f}{2}\right)-\left(\frac{1-f}{2}\right)\log_2 \left(\frac{1-f}{2}\right)
\end{equation}
of the theoretical and experimental qubit states against coupling strength, where $f$ was defined in Eqn.~\ref{general state}. Theoretical curves were calculated using equation \ref{form B} for $f_B$ and $f_C=1-f_B$. The qubit B is the system qubit for the pure-to-mixed homogenisation, while C is the system qubit for the mixed-to-pure homogenisation. As expected, the qubit being transformed from a mixed to a pure state decreases in entropy, with the effect being strongest for strong coupling, while the qubit being transformed from a pure to a mixed state increases in entropy. Since all the interactions are unitary, the total von Neumann entropy of the combined homogeniser and system must remain constant. Hence it can be deduced that the entropy decrease of the mixed-to-pure system qubit C must be accompanied by an increase in the entropy of the homogeniser (qubits A and B), which is the irreducible entropic cost associated with preparing a pure state.

\begin{figure}[tb]
\centering
\includegraphics[width=85mm]{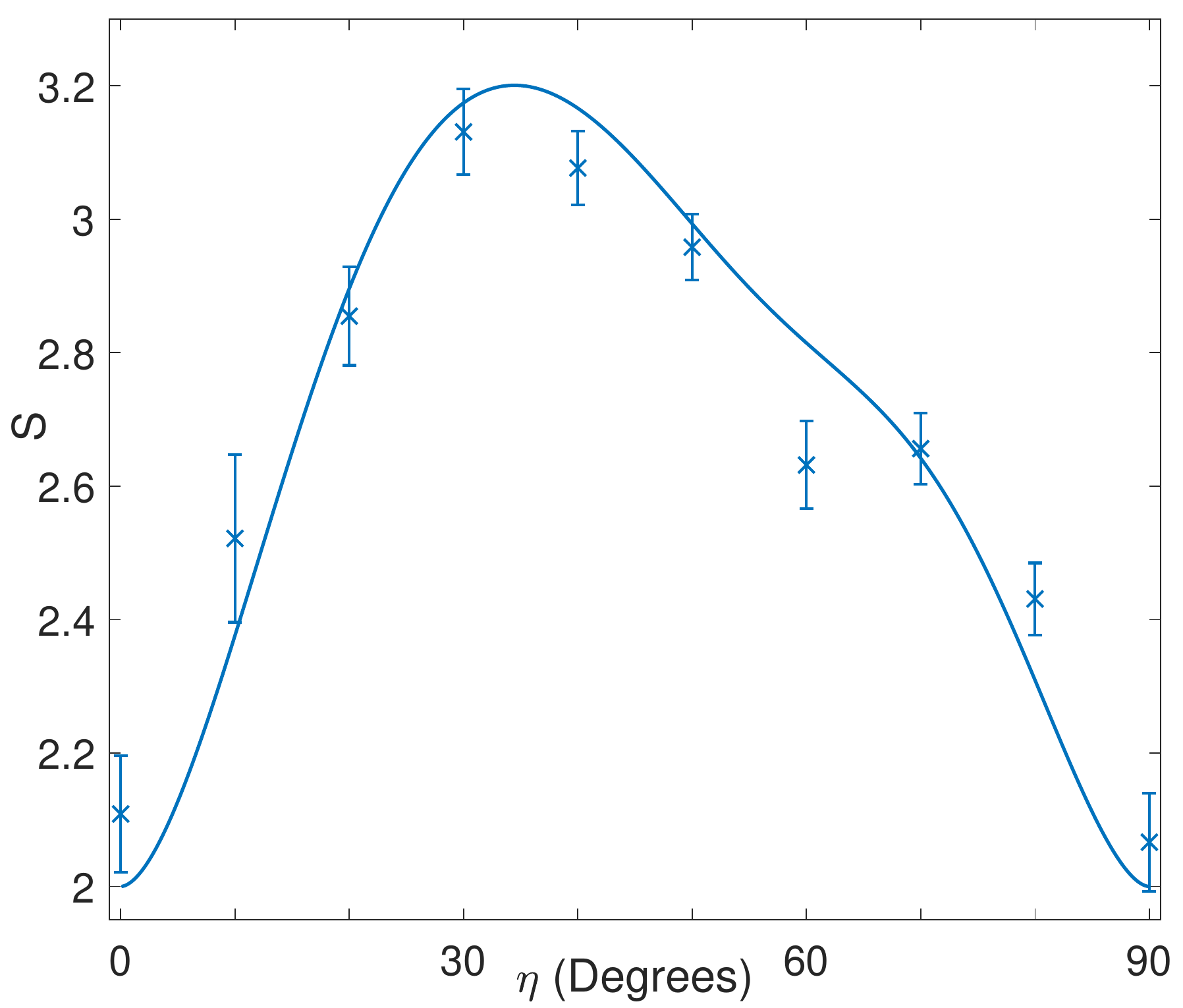}
\caption{Total von Neumann entropies of the four individual qubits, calculated from the experimental results and plotted against the theoretical predictions.}
\label{Total VN}
\end{figure}

\paragraph*{Scrambling.---}
The initial pure or mixed state of a system qubit becomes indistinguishable from the original state of the homogeniser qubits, which is a special case of information scrambling. As explained in \cite{ziman_quantum_2001}, the information about the system's initial state becomes hidden in mutual correlations between the homogeniser qubits. If there were no mutual correlations, one would expect the sum of the von Neumann entropies of the four qubits to equal two for all coupling strengths. The actual sum of the von Neumann entropies is in Figure \ref{Total VN}. While this is two for the cases of an identity or a SWAP operation, for intermediate coupling strengths it is larger. This indicates that the negative contribution to von Neumann entropy from mutual correlations has been unaccounted for, which is due to considering only reduced density operators to describe the qubit states.

\paragraph*{Reusability.---}
The results in \cite{ziman_quantum_2001} show that for the regime of weak coupling and a large reservoir, the reservoir qubits remain almost unchanged by the interaction. We can therefore hypothesise that the same homogeniser could be reused for a second time (or more) to successfully homogenise more system qubits. Whilst the system we have tested here only has a small homogeniser, we can explore the hypothesis by comparing the states of the four qubits in the weak coupling regime.

Qubit A can be interpreted as interacting with a homogeniser that has already been ``used" once, to homogenise B. Similarly, qubit D can be interpreted as being homogenised by a homogeniser that has already been used to homogenise C. These second system qubit states are closely aligned with the first system qubit counterparts for weak coupling, and diverge for strong coupling, in figure \ref{experiment}. This indicates that the homogeniser is minimally changed from its original state in the weak coupling regime, allowing it to perform just as effective a homogenisation on the second system qubit as it did on the first. Hence, the homogeniser can be reused to some extent to give the same incremental changes in system qubit state, in the weak coupling regime. Quantifying how far the homogeniser can be reused, depending on its initial state, is a line of further theoretical and experimental work in this area, which we leave for a future paper.

\section{Conclusions}

We have performed an NMR demonstration of the quantum homogeniser, using the partial swap on a system of four qubits. This demonstrates the principle behind a machine that can perform processes such as information erasure and the preparation of pure states using entirely unitary interactions. The experiments show the homogenisation of a pure state and of a mixed state. The asymmetry in the evolution of the states with coupling strength can be explained by the decoherence within the experiment. After accounting for this, the results are consistent with the theoretical symmetric evolution of the pure and mixed qubit states.

Our experiment was limited to showing the principles behind a homogenisation machine with a small number of NMR qubits. In theory, the quantum homogeniser can perform homogenisation to arbitrarily high accuracy in the limit of an infinitely large reservoir, with vanishing deterioration to its own state in the limit of weak coupling. This is of particular interest for a number of approaches which aim to study entities that undergo no net change while enabling a transformation: resource theory \cite{Coeke2014}, where such entities are called ‘catalysts’, and the theory of quantum reference frames \cite{Popescu2018}, in the studies of the emergence of classicality. It is also of particular interest for the constructor theory of thermodynamics \cite{deutsch_constructor_2014, marletto_constructor_2017}, where thermodynamic irreversibility is associated with there being tasks that can be performed to arbitrarily high accuracy in one direction, but not in the reverse. The homogeniser is therefore a candidate toy-model to demonstrate these phenomena within quantum theory.

Expanding the experimental demonstration of the homogeniser to different regimes and quantum technologies could give an additional insight into the fundamental limits to homogenising pure and mixed states by physical machines. This may ultimately help define the limiting capabilities of quantum computation protocols, and new devices in the field of quantum thermodynamics.

\begin{acknowledgments}
GB is supported by a Felix Scholarship. CM thanks the Templeton World Charity Foundation and the Eutopia Foundation. VV's research is supported by the National Research Foundation, Prime Minister's Office, Singapore, under its Competitive Research Programme (CRP Award No. NRF- CRP14-2014-02) and administered by Centre for Quantum Technologies, National University of Singapore. Quantum circuits were drawn using Quantikz \cite{Kay2018}. This publication was made possible through the support of the ID 61466 grant from the John Templeton Foundation, as part of the The Quantum Information Structure of Spacetime (QISS) Project (qiss.fr). The opinions expressed in this publication are those of the authors and do not necessarily reflect the views of the John Templeton Foundation. This research was also supported by grant number (FQXi-RFP-1812) from the Foundational Questions Institute and Fetzer Franklin Fund, a donor advised fund of Silicon Valley Community Foundation.
\end{acknowledgments}

\appendix
\section{The NMR spin system}\label{NMR}
The NMR sample comprises a solution of fully \nuc{13}{C} labelled crotonic acid (\textit{trans}-$\rm CH_3 CH\!=\!CH CO_2H$) dissolved in deuterated acetone and held at a temperature of 300\,K. The structure is shown in Fig.~\ref{ham}) as a ``ball and stick'' diagram, with carbon atoms coloured black, hydrogen atoms white, and oxygen atoms red. As usual for liquid state NMR samples rapid molecular tumbling means that the different molecules are rendered effectively identical, and direct dipole--dipole couplings between nuclei in different molecules are averaged out, so the system can be treated as an ensemble of identical independent copies. The deuterium signal of the solvent is used to provide a field-frequency lock, which holds the NMR transition frequencies constant. 

The spin system contains ten spin-$1/2$ entities, four \nuc{13}{C} nuclei and six \nuc{1}{H} nuclei. The two \nuc{16}{O} nuclei can be completely ignored as they are spin-0, and so have no effect on the other spins. The \nuc{1}{H} nucleus in the hydroxyl group can also be ignored as it undergoes rapid chemical exchange \cite{EBWbook}, averaging out its interaction with other spins, leaving us with a nine spin system. This can be described by the Hamiltonian
\begin{equation}
\ham = \ham_{\mathrm{H}} + \ham_{\mathrm{C}} + \ham_{\mathrm{HC}}
\end{equation}
where $\ham_{\mathrm{H}}$ and $\ham_{\mathrm{C}}$ comprises the chemical shifts of and homonuclear couplings \cite{EBWbook} between the five \nuc{1}{H} and four \nuc{13}{C} nuclei respectively, while $\ham_{\mathrm{HC}}$ represents the heteronuclear couplings between the different nuclear types.

The RF transmitter frequency was set to 150.852078\,MHz, which corresponds to zero-frequency in the NMR spectrum shown in Fig.~\ref{ham}. For this spectrum the $\ham_{\mathrm{HC}}$ terms were removed by \nuc{1}{H} decoupling \cite{EBWbook}. The frequencies in $ \ham_{\mathrm{C}}$ are given in the table below.
\begin{center}
\begin{tabular}{|c|c|c|c|c|}
\hline
\makebox[14mm]{A}&\makebox[14mm]{B}&\makebox[14mm]{C}&\makebox[14mm]{D}&\makebox[14mm]{}\\\hline
-11962.2&41.6&1.5&7.1&A  \\\hline
&7306.0&69.6&1.2&B\\\hline
&&3972.1&72.3&C\\\hline
&&&10626.1&D\\\hline
\end{tabular}
\end{center}
Here the resonance frequencies for each spin are given down the diagonal, and couplings between spins are listed off the diagonal; all frequencies are in Hz.

\section{Robust GRAPE pulses}\label{Robust}
Of the nine relevant spin-$1/2$ nuclei in the spin system (see Appendix~\ref{NMR}), we only seek to perform quantum gates on the four \nuc{13}{C} nuclei, which form the system qubits, with the five \nuc{1}{H} nuclei providing an unwanted environment. The traditional approach is to apply a broadband decoupling sequence \cite{EBWbook} to the \nuc{1}{H} environment qubits so as to average out their interactions with the system qubits. As the environment spins are not directly observed, decoupling the environment qubits is equivalent to deleting the $\ham_{\mathrm{H}}$ and $\ham_{\mathrm{HC}}$ terms from the overall Hamiltonian thereby leaving just the Hamiltonian $\ham_{\mathrm{C}}$, and GRAPE \cite{Khaneja2005} can now be used to design quantum gates on this four-qubit system.

However, in practice it is not possible to achieve perfect \nuc{1}{H} decoupling, completely tracing out the environment qubits so as to allow an ideal four-qubit treatment of the system. Perfect decoupling requires extremely high RF powers, beyond realistic hardware capabilities, and will also led to heating of the sample. Our simulations suggest that imperfect \nuc{1}{H} decoupling is the main source of error encountered while implementing GRAPE pulses on this system. The most direct approach to overcome this is to design a nine-qubit quantum gate that controls both \nuc{1}{H} and \nuc{13}{C} excitation so as to perform an identity operation on the environment qubits, while implementing the desired four-qubit operation on system qubits. However, designing nine-qubit gates using GRAPE is computationally expensive and impractical.

Here, we adopt the approach described in Ref.\ \cite{Bhole2020} for designing GRAPE pulses that totally avoid environment decoupling and also have a similar computational complexity scaling to four-qubit systems. The idea is to leave the \nuc{1}{H} nuclei completely untouched while implementing the quantum gate, so that they can be treated as remaining in fixed eigenstates. These five spins give rise to $2^5=32$ possible eigenstates $\{00000, 00001 \dots 11111\}$, but there are only $16$ genuinely distinct eigenstates as the three \nuc{1}{H} nuclei in the methyl group are completely equivalent. By considering each of these $16$ eigenstates one at a time, we obtain $16$ separate system Hamiltonians. Designing a GRAPE pulse that is robust over these $16$ Hamiltonians is equivalent to tracing out $\ham_{\mathrm{H}}$ and $\ham_{\mathrm{HC}}$ terms from the combined Hamiltonian. Therefore, at modest computational overhead, the GRAPE pulses can be made robust to the environment qubits.

\section{Phase-only GRAPE pulses}\label{PhaseOnly}
In GRAPE \cite{Khaneja2005}, the control sequence is made piece-wise continuous by discretizing the total control duration $T$ into $N$ segments each of duration $\Delta t = T/N$. Generally, the controls for each segment $j$ are characterized by an amplitude $\Omega(j)$ and phase $\phi(j)$ which are constant for a duration $\Delta t$. However, here we restrict the control sequence to have a fixed amplitude $\Omega$ across all $N$ segments. In such a case, the propagator during the $j^{\mathrm{th}}$ segment for a homonuclear spin system is given by
\begin{equation}
U_j = \exp\left(-\frac{\I}{\hbar} \left[\ham_0 + \Omega\ \cos{\phi(j)} I_x + \Omega\ \sin{\phi(j)} I_y  \right] \Delta t \,\right)\label{U_j}
\end{equation}
where $\ham_0$ is the internal Hamiltonian of the spin system and $I_x$ and $I_y$ are the total $x$- and $y$-Pauli spin-$1/2$ operators acting on all the spins. As described in Ref.\ \cite{Bhole2018}, this propagator can be expressed as a series of $z$- and $x$-rotations,
\begin{equation}
U_j = \mathcal{Z}_j\mathcal{X}\mathcal{Z}_j^{\dagger}
\end{equation}
where, $\mathcal{Z}_j = \exp{[-\I\phi(j) I_z]}$ is a diagonal matrix and $\mathcal{X} = \exp{[-\I(\ham_0 + \Omega I_x)\Delta t]}$ is a constant matrix, which is the same for all $N$ segments. The operator $\mathcal{X}$ can thus be evaluated once, and then stored for reuse, while the diagonal matrix $\mathcal{Z}_j$ can be treated as a column vector to perform element-wise multiplication with $\mathcal{X}$. As a result, the evaluation of the propagator $U_j$ can be greatly sped up, with the only computation required being element-wise multiplication of a matrix and a vector. Thus, by setting a fixed amplitude across all segments, it is possible to entirely avoid the expensive computation of matrix exponentials, while simultaneously avoiding the need for amplitude penalty functions.

Further, this construction of the propagator can greatly simplify the evaluation of gradient of the propagator with respect to the control variables, a necessary step for the GRAPE algorithm to update the controls. The only variable controls present in the propagator $U_j$ are the phases $\phi(j)$ that appear in the diagonal matrix $\mathcal{Z}_j$. Since, $\mathcal{Z}_j$ is diagonal, it is also possible to evaluate the exact result
\begin{equation}
{\rd\mathcal{Z}_j}/{\rd\phi(j)}= -\I I_z \mathcal{Z}_j
\end{equation}
analytically. Applying the chain rule, the exact gradient of the propagator $U_j$ with respect to the phase $\phi_j$ is
\begin{equation}
\frac{\rd U_j}{ \rd\phi(j)} = \I [U_j, I_z].
\end{equation}
Unlike the original GRAPE algorithm \cite{Khaneja2005} which relied on gradients approximated to first-order, our method can evaluate exact analytic gradients at no additional cost.

Methods for evaluating exact gradients have been discussed in \cite{Machnes2011}, where evaluation of the propagators by eigendecomposition gives the exact gradients at no additional cost. However, in our method, by simply fixing the amplitudes the exact gradients can be calculated without the need of matrix exponentiation or eigendecomposition for evaluating propagators. Moreover, exact gradients are necessary \cite{Jensen2020} while using second order optimization routines like BFGS \cite{DeFouquieres2011}. Further, the exact Hessian can also be evaluated analytically if desired, giving accelerated convergence while using Netwon--Raphson type methods \cite{Goodwin2016}.

\bibliography{qHomo}
\end{document}